# SOME SIMPLIFICATIONS FOR THE EXPECTATION-MAXIMIZATION (EM) ALGORITHM: THE LINEAR REGRESSION MODEL CASE

Daniel A. Griffith
Ashbel Smith Professor (Emeritus, as of 2024)
School of Economic, Political and Policy Sciences
University of Texas at Dallas
800 W. Campbell Rd., Richardson, TX 75080-3021
Phone: *****; Fax: *****
e-mail: dagriffith@utdallas.edu

ABSTRACT The EM algorithm is a generic tool that offers maximum likelihood solutions when datasets are incomplete with data values missing at random or completely at random. At least for its simplest form, the algorithm can be rewritten in terms of an ANCOVA regression specification. This formulation allows several analytical results to be derived that permit the EM algorithm solution to be expressed in terms of new observation predictions and their variances. Implementations can be made with a linear regression or a nonlinear regression model routine, allowing missing value imputations, even when they must satisfy constraints.

Fourteen example datasets gleaned from the EM algorithm literature are reanalyzed. Imputation results have been verified with SAS PROC MI. Six theorems are proved that broadly contextualize imputation findings in terms of the theory, methodology, and practice of statistical science.

## 1. MOTIVATION

The *E*xpectation-*M*aximization (EM) algorithm [3], an iterative procedure for computing maximum likelihood estimates (MLEs) when datasets are incomplete, with data values missing at random (MAR) or completely at random (MCAR), is a useful device for helping to solve a wide range of model-based estimation problems[1]. Flury and Zoppè [7, p. 209] emphasize that

> it can not be stressed enough that the E-step does not simply involve replacing missing data by their conditional expectations (although this is true for many important applications of the algorithm).

But frequently model-based estimation problems desire just this type of imputation output from the algorithm, especially for small area estimation problems [e.g., 15, 16]. Furthermore, in certain situations, focusing on imputation dramatically simplifies the EM solution.

More recent descriptions of the EM algorithm may be found in Flury and Zoppè [7], Meng [13], and McLachlan and Krishnan [12], among others. The objective of this paper is to present regression solutions that render conditional expectations for missing values in a dataset that are equivalent to EM algorithm results when: values are missing only in the response variable, Y; values are missing on both sides of a regression equation; and, missing values are constrained in some way (e.g., sum to a given total). To this end, a formulation of the EM algorithm in terms of a linear regression problem is presented in order to add insight about imputations and their variances. Because the EM procedure requires imputation of the complete-data sufficient statistics, rather than just the individual missing observations, the equivalency discussed here initially derives from an assumption of normality, for which the means and covariances constitute the sufficient statistics. An assumption of normality links ordinary least squares (OLS) and maximum likelihood (ML) estimation regression results, too; application of the Rao-Blackwell factorization theorem verifies that the means and covariances are sufficient statistics in this situation.

---
[1] Qui et al. (2000), for example, address non-MCAR cases.

## 2. BACKGROUND

Yates [21] shows for analysis of variance (ANOVA) that if each missing observation is replaced by a parameter to be estimated (i.e., the conditional expectation for a missing value), the resulting modified analysis becomes straightforward by treating the estimated missing value as a parameter to be estimated (i.e., an imputation). Rewriting the ANOVA as an OLS regression problem involves introducing a binary indicator variable for each missing value—the value of 1 denoting the missing value observation in question, and 0 otherwise—with the estimated regression coefficients for these indicator variables being the negative of the missing value estimates. This approach is equivalent to subtracting each observation's missing value, in turn, from each side of a regression equation. Generalizing this regression formulation to include covariates allows missing values to be estimated with an analysis of covariance (ANCOVA) regression specification, one in fact suggested by Bartlett [2] and by Rubin [18], and more extensively discussed in Little and Rubin [11, pp. 30-39]. Replacing the arbitrarily assigned value of 1 in each individual observation missing value indicator variable by the value -1 yields estimated regression parameters with the correct sign.

Consider a bivariate set of n observed values, each pair denoted by ($y_i$, $x_i$), i=1, 2, ..., n. Suppose only the response variable, Y, contains incomplete data [also see 10], which are missing completely at random (i.e., the mechanism creating missing values depends upon neither the observed nor the unobserved data). First, the $n_m$ missing values need to be replaced by 0. Second, $n_m$ 0/-1 indicator variables, $-I_m$ (m = 1, 2, ..., $n_m$), need to be constructed; $I_m$ contains n-1 0s and a single 1 corresponding to the $m^{th}$ missing value observation. The minus sign for $–I_m$ indicates that a -1 actually is entered into each of the m indicator variables. Regressing Y on a complete data predictor variable, X—which furnishes the redundant information that is exploited to compute imputations—together with the set of m indictor variables constitutes the ANCOVA.

Suppose $\mathbf{Y}_o$ denotes the $n_o$-by-1 ($n_o$ = n – $n_m$) vector of observed response values, and $\mathbf{Y}_m$ denotes the $n_m$-by-1 vector of missing response values. Let $\mathbf{X}_o$ denote the vector of predictor values for the set of observed response values, and $\mathbf{X}_m$ denote the vector of predictor values for the set of missing response values. Further, let **1** denote an n-by-1 vector of ones that can be partitioned into $\mathbf{1}_o$, denoting the vector of ones for the set of observed response values, and $\mathbf{1}_m$, denoting the vector of ones for the set of missing response values. Then the ANCOVA specification of the regression model may be written in partitioned matrix form as

$$\begin{pmatrix} \mathbf{Y}_o \\ \mathbf{0}_m \end{pmatrix} = \begin{pmatrix} \mathbf{1}_o & \mathbf{X}_o \\ \mathbf{1}_m & \mathbf{X}_m \end{pmatrix} \begin{pmatrix} \beta_0 \\ \beta_X \end{pmatrix} + \begin{pmatrix} \mathbf{0}_{om} \\ -\mathbf{I}_{mm} \end{pmatrix} (\boldsymbol{\beta}_m) + \begin{pmatrix} \boldsymbol{\varepsilon}_o \\ \mathbf{0}_m \end{pmatrix}, \qquad (1)$$

where $\mathbf{0}_j$ (j = o, m) is an $n_j$-by-1 vector of zeroes, $\mathbf{0}_{om}$ is an $n_o$-by-$n_m$ matrix of zeroes, $\beta_0$ and $\beta_X$ respectively are the bivariate intercept and slope regression parameters, $\boldsymbol{\beta}_m$ is an $n_m$-by-1 vector of conditional expectations expressed as regression parameters, $\mathbf{I}_{mm}$ is an $n_m$-by-$n_m$ identity matrix, and $\boldsymbol{\varepsilon}_o$ is an $n_o$-by-1 vector of random error terms. The bivariate OLS regression coefficient estimates, $b_0$ and $b_X$, of $\beta_0$ and $\beta_X$, respectively, for this ANCOVA specification are given by

$$\begin{pmatrix} b_0 \\ b_X \end{pmatrix} = \begin{pmatrix} n - n_m & \mathbf{1}_o^T \mathbf{X}_o \\ \mathbf{X}_o^T \mathbf{1}_o & \mathbf{X}_o^T \mathbf{X}_o \end{pmatrix}^{-1} \begin{pmatrix} \mathbf{1}_o^T \mathbf{Y}_o \\ \mathbf{X}_o^T \mathbf{Y}_o \end{pmatrix}, \qquad (2)$$

where T denotes matrix transpose, which is the regression result for the observed data only. In addition, the regression coefficients, $\mathbf{b}_m$, for the indicator variables are given by

$$\mathbf{b}_m = b_0 \mathbf{1}_m + b_X \mathbf{X}_m = \hat{\mathbf{Y}}_m, \qquad (3)$$

which is the vector of point estimates for additional observations (i.e., the prediction of new observations) that practitioners recommend should have Y values within the interval defined by the extreme values contained in the vector $\mathbf{Y}_o$. This is a standard OLS regression result, as is the prediction error that can be attached to it [see, for example, 14, pp. 37-39]. Little [10, p. 1233] alludes to this result by noting that "EM starting from [complete case] estimates converges in one iteration." In addition, the values here are positive because the $-\mathbf{I}_m$ indicator variables contain negative ones.

Dodge [4, p. 159] cautions that the OLS equivalency highlighted here rests on the existence of estimable equations, which in some instances means that the ANCOVA solution is appropriate only when the number of missing values is not excessive. If enough observations are missing, the number of degrees of freedom can become zero or negative, the matrix $\begin{pmatrix} n_o & \mathbf{1}_o^T \mathbf{X}_o \\ \mathbf{X}_o^T \mathbf{1}_o & \mathbf{X}_o^T \mathbf{X}_o \end{pmatrix}$ can become singular, and as such not all of the parametric functions would be estimable. But in selected instances, standard regression routines can be tricked into still computing results.

These results may be generalized with ML techniques as follows, where the likelihood function is given by

$$L = \frac{1}{(2\pi)^{n/2} \sigma^n} e^{-\frac{1}{2\sigma^2} \left[ \begin{pmatrix} \mathbf{Y}_0 \\ \mathbf{0}_m \end{pmatrix} - \begin{pmatrix} \mathbf{X}_o & \mathbf{0}_o \\ \mathbf{X}_m & -\mathbf{I}_m \end{pmatrix} \begin{pmatrix} \boldsymbol{\beta} \\ \boldsymbol{\beta}_m \end{pmatrix} \right]^T \left[ \begin{pmatrix} \mathbf{Y}_0 \\ \mathbf{0}_m \end{pmatrix} - \begin{pmatrix} \mathbf{X}_o & \mathbf{0}_o \\ \mathbf{X}_m & -\mathbf{I}_m \end{pmatrix} \begin{pmatrix} \boldsymbol{\beta} \\ \boldsymbol{\beta}_m \end{pmatrix} \right]}. \qquad (4)$$

THEOREM 1. If the errors, $\boldsymbol{\varepsilon} = \mathbf{Y} - \mathbf{X}\boldsymbol{\beta}$, in equation (1) are distributed as $\text{MVN}(\mathbf{0}, \mathbf{I}\sigma^2)$, then $\hat{\boldsymbol{\beta}}_m = \mathbf{X}_m \boldsymbol{\beta}$ is a MLE.

PF: $\dfrac{\partial \text{LN}(L)}{\partial \boldsymbol{\beta}_m} = -\dfrac{1}{2\sigma^2} [2\boldsymbol{\beta}_m - \mathbf{X}_m \boldsymbol{\beta} - \mathbf{X}_m \boldsymbol{\beta}] = \mathbf{0}$

$$\therefore \hat{\boldsymbol{\beta}}_m = \mathbf{X}_m \boldsymbol{\beta}$$

$$\frac{\partial^2 \text{LN(L)}}{\partial \beta_m^2} = -\frac{1}{\sigma^2} \forall m$$

$$\therefore \hat{\boldsymbol{\beta}}_m = \mathbf{X}_m \boldsymbol{\beta} \text{ is a MLE. } \square$$

Because **b**, the estimate of $\boldsymbol{\beta}$, is a function of $\mathbf{Y}_o$, the unknown missing values still follow a normal distribution. Equation (4) is equivalent to the independent and identically distributed case, and can be rewritten as such by not using matrix notation, and separating the observed and missing data density in L. The generalization of this result relates to the missing information principle [15].

Furthermore,

THEOREM 2. If the errors, $\boldsymbol{\varepsilon} = \mathbf{Y} - \mathbf{X}\boldsymbol{\beta}$, in equation (1) are distributed as $\text{MVN}(\mathbf{0}, \mathbf{I}\sigma^2)$, then $\hat{\boldsymbol{\beta}} = (\mathbf{X}_o^T \mathbf{X})^{-1} \mathbf{X}_o^T \mathbf{Y}_o = \mathbf{b}_o$ is a MLE.

PF: $\dfrac{\partial \text{LN(L)}}{\partial \boldsymbol{\beta}} =$

$$-\frac{1}{2\sigma^2}\left[-\mathbf{X}_o^T \mathbf{Y}_o - \mathbf{X}_o^T \mathbf{Y}_o + 2\mathbf{X}_o^T \mathbf{X}_o \boldsymbol{\beta} - \mathbf{X}_m^T \mathbf{Y}_m - \mathbf{X}_m^T \mathbf{Y}_m + 2\mathbf{X}_m^T \mathbf{X}_m \boldsymbol{\beta}\right] = \mathbf{0}$$

$$\mathbf{X}_o^T \mathbf{X}_o \boldsymbol{\beta} + \mathbf{X}_m^T \mathbf{X}_m \boldsymbol{\beta} = \mathbf{X}_o^T \mathbf{Y}_o + \mathbf{X}_m^T \mathbf{Y}_m = \mathbf{X}_o^T \mathbf{Y}_o + \mathbf{X}_m^T \mathbf{X}_m \boldsymbol{\beta}$$

$$\therefore \boldsymbol{\beta} = (\mathbf{X}_o^T \mathbf{X}_o)^{-1} \mathbf{X}_o^T \mathbf{Y}_o = \mathbf{b}_o$$

$$\frac{\partial \left[\frac{\partial \text{LN(L)}}{\partial \boldsymbol{\beta}}\right]}{\partial \boldsymbol{\beta}} = -\frac{1}{2\sigma^2}\left(2\mathbf{X}_o^T \mathbf{X}_o + 2\mathbf{X}_m^T \mathbf{X}_m\right), \text{ which is negative definite}$$

$$\therefore \boldsymbol{\beta} = (\mathbf{X}_o^T \mathbf{X}_o)^{-1} \mathbf{X}_o^T \mathbf{Y}_o = \mathbf{b}_o \text{ is a MLE. } \square$$

In the case of bivariate linear regression, $\mathbf{b}_o = \begin{pmatrix} b_0 \\ b_X \end{pmatrix}$. In addition, these results are the same as those obtained with OLS, which is demonstrated in the ensuing analyses.

## 2.1. An iterative example

Consider the example in McLachlan and Krishnan [12, pp. 49-51] for which iterative sufficient statistics results are reported. Bivariate linear regression theory states that

$$b_X = \frac{\hat{\sigma}_{XY}}{\hat{\sigma}_X^2} \quad \text{and} \quad b_0 = \hat{\mu}_Y - b_X \hat{\mu}_X .$$

Computing $b_0$ and $b_X$ from the tabulated iterative numerical values reported in McLachlan and Krishnan render the corresponding results appearing in Table 1.

Table 1. Iterative results for a bivariate data example presented in McLachlan and Krishnan [12, pp. 49-51].

| Iteration | McLachlan and Krishnan | | | Nonlinear regression results | | |
|---|---|---|---|---|---|---|
| | $b_X$ | $b_0$ | $-2\log(L)$ | $b_X$ | $b_0$ | Error sum of squares |
| 0 | 0.80535 | 4.40547 | 1019.64200 | 0.8054 | 4.4055 | 159.9 |
| 1 | 0.52832 | 7.68467 | 210.93090 | 0.5283 | 7.6847 | 127.3 |
| 2 | 0.52050 | 7.83272 | 193.33120 | 0.5205 | 7.8327 | 127.2 |
| 3 | 0.51973 | 7.85517 | 190.55040 | 0.5197 | 7.8552 | 127.2 |
| 4 | 0.51957 | 7.85997 | 190.01470 | 0.5196 | 7.8600 | 127.2 |
| 5 | 0.51954 | 7.86104 | 189.90800 | 0.5195 | 7.8610 | 127.2 |
| 6 | 0.51953 | 7.86125 | 189.88660 | 0.5195 | 7.8613 | 127.2 |
| 7 | 0.51953 | 7.86133 | 189.88230 | | | |
| 8 | 0.51953 | 7.86133 | 189.88160 | | | |
| 9 | 0.51953 | 7.86132 | 189.88140 | | | |
| 10 | 0.51953 | 7.86132 | 189.88140 | | | |
| 11 | 0.51953 | 7.86132 | 189.88140 | | | |

The ANCOVA form of the linear regression problem can be formulated as the following nonlinear regression problem, which is the well-known EM algorithm, where $\tau$ is the iteration counter:

Step 1: initialize $b_{0,\tau=0}$ and $b_{X,\tau=0}$

Step 2: $\hat{\mathbf{Y}}_{m,\tau} = b_{0,\tau}\mathbf{1}_m + b_{X,\tau}\mathbf{X}_m$, and hence $\mathbf{Y}_\tau = \begin{pmatrix} \mathbf{Y}_o \\ \hat{\mathbf{Y}}_{m,\tau} \end{pmatrix}$

Step 3: regress $\mathbf{Y}_\tau$ on X to obtain iteration estimates $a_{\tau+1}$ and $b_{\tau+1}$

Step 4: repeat Steps 2 and 3, replacing the $\tau$ with $\tau+1$ results in each iteration, until $a_{\tau+1}$ and $b_{\tau+1}$, and hence $\hat{\mathbf{Y}}_{m,\tau}$, converge.

This iterative procedure can be implemented with a nonlinear regression routine (e.g., SAS PROC NLIN). Beginning with the same initial values (i.e., iteration 0) used by McLachlan and Krishnan for their problem renders the corresponding results appearing in Table 1.

Two points merit discussion here. First, if $b_{X,\tau=0} = b_X$ and $b_{0,\tau=0} = b_0$, respectively the slope and the intercept estimates obtained by regressing only the observed values of Y on their corresponding values of X, then a nonlinear regression routine instantly converges (i.e., for $\tau = 0$) with $b_{0,\tau=0}$ and $b_{X,\tau=0}$. Second, with alternative initial values, the nonlinear regression routine converges faster than the conventional EM algorithm (e.g., 7 rather than 11 iterations in Table 1). These findings suggest the following theorem:

THEOREM 3. When missing values occur only in a response variable, Y, then the iterative solution to the EM algorithm produces the regression coefficients, $\mathbf{b}_o$, calculated with only the complete data.

PF: Let **b** denote the vector of regression coefficients that is converged upon. Then if $\hat{\mathbf{Y}}_m = \mathbf{X}_m\mathbf{b}$,

$$\mathbf{b} = \left[\begin{pmatrix}\mathbf{X}_o\\\mathbf{X}_m\end{pmatrix}^T\begin{pmatrix}\mathbf{X}_o\\\mathbf{X}_m\end{pmatrix}\right]^{-1}\begin{pmatrix}\mathbf{X}_o\\\mathbf{X}_m\end{pmatrix}^T\begin{pmatrix}\mathbf{Y}_o\\\mathbf{X}_m\mathbf{b}\end{pmatrix}$$

$$= (\mathbf{X}_o^T\mathbf{X}_o + \mathbf{X}_m^T\mathbf{X}_m)^{-1}(\mathbf{X}_o^T\mathbf{Y}_o + \mathbf{X}_m^T\mathbf{X}_m\mathbf{b})$$

$$= (\mathbf{X}_o^T\mathbf{X}_o)^{-1}\mathbf{X}_o^T\mathbf{Y}_o = \mathbf{b}_o$$

∴ **b** is equivalent to the vector of known data regression coefficients, which yields the point estimate prediction equation for a new observation. □

This equivalency of regression coefficients, which also is noted by Little and Rubin [11, p. 237], and also is an MLE (by Theorem 2), is demonstrated in Figures 1a and 1b, for which 10,000 multiple imputations were computed using SAS PROC MI [see 8] for the set of examples appearing in Table 2. This comparison verifies that the closed-form and the EM results are the same, and also furnishes a simplified method for computational purposes. This finding furnishes a simplified version of the EM algorithm in this case. In addition, for the example presented by Little and Rubin [11, p. 138] for which explicit MLE formulae exist and that is analyzed in McLachlan and Krishnan [12, p. 49],

$\hat{w}_{2,9} = 12.53711$ and $\hat{w}_{2,10} = 14.61523$, rendering

$$\left(\sum_{i=1}^{8} w_{i,2} + \hat{w}_{9,2} + \hat{w}_{10,2}\right)/10 = 14.61523,$$

which is equivalent to $\hat{\mu}_2 = 119/8 + (199.5/384)^2(130/10 - 108/8)$,

$$s_{11} = \sum_{i=1}^{8}(w_{i,1} - \frac{108}{8})^2/8 \ , s_{22} = \sum_{i=1}^{8}(w_{i,2} - \frac{119}{8})^2/8 \ , \text{and}$$

$$s_{12} = \sum_{i=1}^{8}(w_{i,1} - \frac{108}{8})(w_{i,2} - \frac{119}{8})/8,$$

which yields $\hat{\sigma}_{22} = 28.85937 + (199.5/384)(402/10 - 384/8) = 26.75405$, and

$$\hat{\sigma}_{12} = [\sum_{i=1}^{8}(w_{i,1} - 13)(w_{i,2} - 14.61523) +$$

$$(9-13)(12.53711 - 14.61523) + (13-13)(15.61523 - 14.61523]/10 = 20.88516.$$

These calculations are those reported in Table 2, the values upon which the EM algorithm iteratively converges [see 12, Table 2.1, p. 50].

### 2.2. Simplicity from equation (1)

Consider a selection of the example datasets contained in Montgomery, Peck and Vining [14], Little and Rubin [11], McLachlan and Krishnan [12], and Schafer [20]. Comparative results appear in Table 2. Results obtained with the ANCOVA procedure outlined in this section are equivalent in all bivariate cases. In addition, as is illustrated in Table 2, associated calculations can be obtained from the ANCOVA results [see 12, p. 49], and the basics for multiple imputation calculations can be established with the ANCOVA results [see 20, p. 195].

The regression result stated in Theorem 3 leads to the following consequence derived from equation (2), which states that imputations together with their standard errors can be computed directly and without resorting to iterative estimation (i.e., a closed-form solution exists):

| Table 2. Calculations from ANCOVA regression and the EM algorithm ||||
|---|---|---|---|
| Data Source | quantity | Reported value | OLS/NLS estimate |
| *McLachlan & Krishnan [12]* | | | |
| p. 49 | $\hat{\mu}_2$ | 14.61523 | 14.61523 |
| | $\hat{\sigma}_{12}$ | 20.88516 | 208.85156/10 |
| | $\hat{\sigma}_{22}$ | 26.75405 | $230.875/8 + 0.51953^2(402/10 - 384/8) = 26.75407$ |
| p. 53 | $\hat{y}(-1,-1)$ | 429.6978 | 429.69767 |
| | $\hat{y}(0,-1)$ | 324.0233 | 324.02326 |
| p. 54 | $\hat{y}_{23}$ | 4.73 | 4.73030 |
| | $\hat{y}_{51}$ | 3.598 | 3.59697 |
| p. 91 | saddle point: $\sigma_{11}, \sigma_{22}, \rho$ | 5/2, 5/2, 0 | 5/2, 5/2, 0 |
| | maxima | 8/3, 8/3, ±0.5 | 2.87977, 2.87977, ±0.88817 |
| *Little & Rubin [11]* | | | |
| p. 34 | $u_1$ estimate | 7.8549 | 7.85492 |
| | $u_2$ estimate | 7.9206 | 7.92063 |
| p. 138 | $\hat{\mu}_2$ | 49.3333 | 49.33333 |
| p. 154 | $\hat{\mu}(x_1)$ | 6.655 | 6.65518 |
| | $\hat{\mu}(x_2)$ | 49.965 | 49.96524 |
| | $\hat{\mu}(x_4)$ | 27.047 | 27.03739 |
| *Schafer [20]* | | | |
| p. 43 | $\mu^{(\infty)}$ | 48.1000 | 48.10000 |
| | $\psi^{(\infty)}$ | 59.4260 | 59.42600 |
| p. 54 | $\hat{\sigma}_{11} = \hat{\sigma}_{22}$ | 1.80 | 18/10 |
| | $\hat{\rho}$ | -1 | -1 |
| | $\hat{\mu}_1 = \hat{\mu}_2$ | 0 | 0 |
| p. 195 | $\hat{y}_{3,2}$ average (n=5) | 226.2 | 228.0 (se = 32.86) |
| | $\hat{y}_{3,4}$ average (n=5) | 146.8 | 146.2 (se = 38.37) |
| | $\hat{y}_{3,5}$ average (n=5) | 190.8 | 192.5 (se = 34.11) |
| | $\hat{y}_{3,10}$ average (n=5) | 250.2 | 271.7 (se = 36.20) |
| | $\hat{y}_{3,13}$ average (n=5) | 234.2 | 241.3 (se = 35.18) |
| | $\hat{y}_{3,16}$ average (n=5) | 269.2 | 269.9 (se = 34.53) |
| | $\hat{y}_{3,18}$ average (n=5) | 192.4 | 201.9 (se = 32.91) |
| | $\hat{y}_{3,23}$ average (n=5) | 215.6 | 207.4 (se = 33.09) |
| | $\hat{y}_{3,25}$ average (n=5) | 250.0 | 255.7 (se = 33.39) |
| *Montgomery, Peck & Vining [14]* | | | |
| pp. 76 & 111 | $\hat{y}_{26}$ | 19.22 | 19.22432 |
| | $s_{\hat{y}_{26}}$ | 3.34628 | $\sqrt{10.6239(1+0.05346)} = 3.34542$ |
| NOTE: SAS does not always produce the same number of digits that are reported in published work. ||||

THEOREM 4. When missing values occur only in a response variable, Y, then by replacing the missing values with zeroes and introducing a binary 0/-1 indicator variable covariate -$I_m$ for each missing value m, such that $I_m$ is 0 for all but missing value observation m and 1 for missing value observation m, the estimated regression coefficient $b_m$ is equivalent to the point estimate for a new observation, and as such furnishes EM algorithm imputations.

PF: Let **$b_m$** denote the vector of regression coefficients for the missing values, and partition the data matrices such that

$$\begin{pmatrix} \mathbf{b}_o \\ \mathbf{b}_m \end{pmatrix} = \left[ \begin{pmatrix} \mathbf{X}_o & \mathbf{0}_{om} \\ \mathbf{X}_m & -\mathbf{I}_{mm} \end{pmatrix}^T \begin{pmatrix} \mathbf{X}_o & \mathbf{0}_{om} \\ \mathbf{X}_m & -\mathbf{I}_{mm} \end{pmatrix} \right]^{-1} \begin{pmatrix} \mathbf{X}_o & \mathbf{0}_{om} \\ \mathbf{X}_m & -\mathbf{I}_{mm} \end{pmatrix}^T \begin{pmatrix} \mathbf{Y}_o \\ \mathbf{0}_m \end{pmatrix}$$

$$= \begin{pmatrix} (\mathbf{X}_o^T \mathbf{X}_o)^{-1} & (\mathbf{X}_o^T \mathbf{X}_o)^{-1} \mathbf{X}_m^T \\ \mathbf{X}_m(\mathbf{X}_o^T \mathbf{X}_o)^{-1} & \mathbf{I}_{mm} + \mathbf{X}_m(\mathbf{X}_o^T \mathbf{X}_o)^{-1} \mathbf{X}_m^T \end{pmatrix} \begin{pmatrix} \mathbf{X}_o^T \mathbf{Y}_o \\ \mathbf{0}_m \end{pmatrix}$$

∴ $\mathbf{b}_o = (\mathbf{X}_o^T \mathbf{X}_o)^{-1} \mathbf{X}_o^T \mathbf{Y}_o$, and

$\mathbf{b}_m = \mathbf{X}_m \mathbf{b}_o$ ,

where $\mathbf{I}_{mm}$ is an $n_m$-by-$n_m$ identity matrix and $\mathbf{0}_{om}$ is a $n_o$-by-$n_m$ null matrix. □

Therefore, **$b_m$** is both an MLE (by Theorems 1 and 2) and the point estimate prediction equation for new observations [see 14, p. 108]. Besides simplicity, one advantage of this result is that the correct number of degrees of freedom is produced by the regression formulation, which actually is an ANCOVA specification, again furnishing a closed-form result for computational purposes. This result also is noted by Little and Rubin [11, p. 238], but without reference to the prediction of new observations. The imputation equivalency is demonstrated in Figure 1c, for which 10,000 multiple imputations were computed using SAS PROC MI for each dataset in the collection of examples appearing in Table 2.

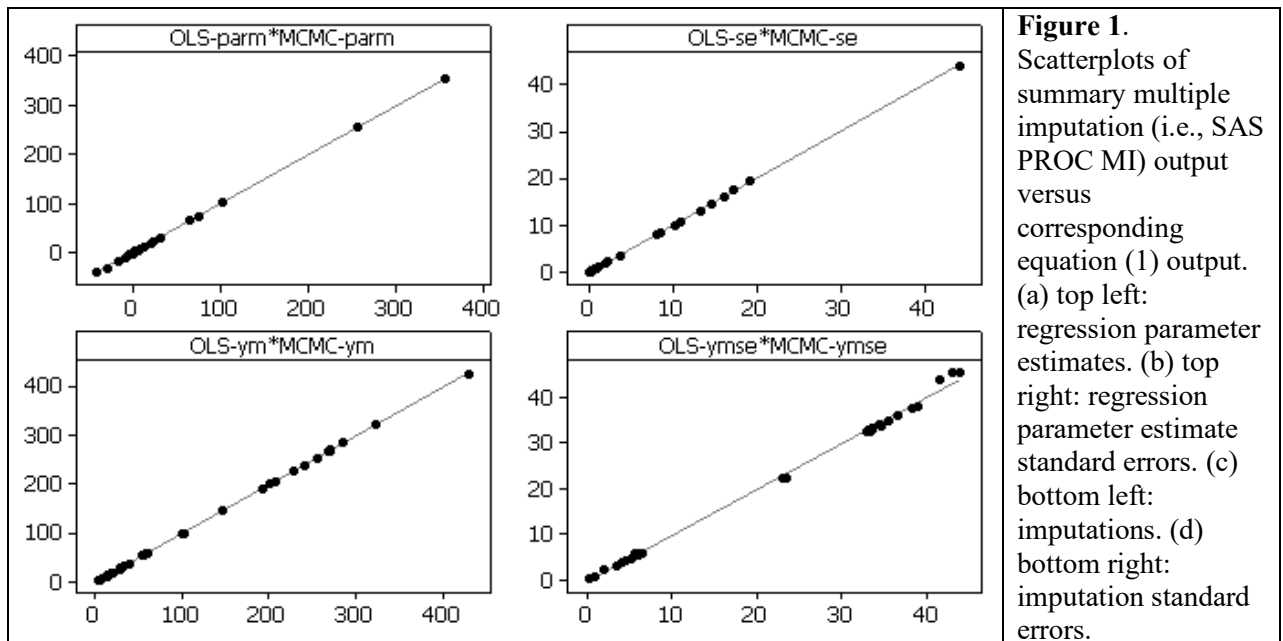

**Figure 1**. Scatterplots of summary multiple imputation (i.e., SAS PROC MI) output versus corresponding equation (1) output. (a) top left: regression parameter estimates. (b) top right: regression parameter estimate standard errors. (c) bottom left: imputations. (d) bottom right: imputation standard errors.

Part of the issue emphasized in arguments such as those by Flury and Zoppè [7], and promoted by the multiple imputation perspective [see 20], is that an imputation is of less value when its variance is unknown. Fortunately, the regression result stated in Theorem 4 leads to the following consequence:

> THEOREM 5. For imputations computed based upon Theorem 4, each standard error of the estimated regression coefficients $\mathbf{b}_m$ is equivalent to the conventional standard deviation used to construct a prediction interval for a new observation, and as such furnishes the corresponding EM algorithm imputation standard error.
>
> PF: Let the subscript *diag* denote diagonal entries, and $\mathbf{s}_{\mathbf{b}_m}$ denote the $n_m$-by-1 vector of imputation standard errors. From standard OLS regression theory, the diagonal elements of the following partitioned matrix furnish the variances of the imputations calculated as regression coefficients:
>
> $$\left[\begin{pmatrix} \mathbf{X}_o & \mathbf{0}_{om} \\ \mathbf{X}_m & -\mathbf{I}_{mm} \end{pmatrix}^T \begin{pmatrix} \mathbf{X}_o & \mathbf{0}_{om} \\ \mathbf{X}_m & -\mathbf{I}_{mm} \end{pmatrix}\right]^{-1} \hat{\sigma}_\varepsilon^2 = \begin{pmatrix} (\mathbf{X}_o^T\mathbf{X}_o)^{-1} & (\mathbf{X}_o^T\mathbf{X}_o)^{-1}\mathbf{X}_m^T \\ \mathbf{X}_m(\mathbf{X}_o^T\mathbf{X}_o)^{-1} & \mathbf{I}_{mm} + \mathbf{X}_m(\mathbf{X}_o^T\mathbf{X}_o)^{-1}\mathbf{X}_m^T \end{pmatrix} \hat{\sigma}_\varepsilon^2$$
>
> $$\therefore \mathbf{s}_{\mathbf{b}_m} = \sqrt{\left\langle [\mathbf{I}_{mm} + \mathbf{X}_m(\mathbf{X}_o^T\mathbf{X}_o)^{-1}\mathbf{X}_m^T]\right\rangle_{diag} \hat{\sigma}_\varepsilon^2} . \;\square$$

Therefore, $\mathbf{s}_{\mathbf{b}_m}$ is the well-known vector of standard errors for the point estimate prediction equation for new observations [see 14, p. 108], furnishing a simplified result for computational purposes. This result is of particular interest because the EM algorithm does not produce standard errors as a by-product [see 9]. The imputation equivalency is demonstrated in Figure 1d, for which 10,000 multiple imputations were computed using SAS PROC MI for the set of examples appearing in Table 2.

Theorems 4 and 5 provide the analytical mean and variance for an EM algorithm imputation when data are missing only in Y. Multiple imputations can be obtained by sampling with a normal probability model whose mean and variance are given by these two theorems. These theorems formalize the statement by McLachlan and Krishnan [12, p. 52] that "the missing cases do not contribute information [about] the model parameters."

## 2.3. Implications

The preceding theorems demonstrate that the EM algorithm reduces to an OLS (as well as ML) linear regression analysis, when a normal probability model can be attached to the error term, whose model specification is as follows:

$$\begin{pmatrix} \mathbf{Y}_o \\ \mathbf{0}_m \end{pmatrix} = \beta_0 \mathbf{1} + \mathbf{X}\boldsymbol{\beta} - \sum_{m=1}^{M} y_m \mathbf{I}_m + \begin{pmatrix} \boldsymbol{\varepsilon}_o \\ \mathbf{0}_m \end{pmatrix}, \qquad (5)$$

where the intercept term is separated from the attribute covariates as in equation (2). The bivariate linear regression equation comparing equation (5) and assembled EM imputation results reported in Table 2 is as follows:

$$\text{reported} = -0.15609 + 1.01245 \times [\text{equation (1) results}] + e,$$
$$(t = -0.13) \quad (t = 1.66)$$

which has an associated $R^2$ value of 0.998. In other words, the bivariate linear regression null hypotheses here of $\alpha = 0$ (for the intercept) and $\beta = 1$ (for the slope parameter) are not rejected. Slight discrepancies detected here are attributable to noise introduced by Schafer [20, p. 195] into the nine multiple imputation results he reports.

### 3. MISSING VALUES IN BOTH X AND Y: THE MULTIVARIATE NORMAL CASE
Little [10, p. 1234] notes that filling in the missing Xs is a more complex problem. One appealing feature of the ANCOVA specification is that it can be generalized for implementation with nonlinear least squares (NLS) procedures in order to handle this imputation problem. Rather than a bivariate situation having missing values in only the response variable, consider a bivariate situation with missing values in both variables, and a multiple variable situation with missing values in a number of the three or more variables. Again, the assumption is that values are missing completely at random.

Suppose a bivariate case has P values missing for variable Y, and Q values missing for variable X, but each observation has at least a $y_i$ or an $x_i$ measure (i.e., no observation has both values missing). Furthermore, those observations having complete ($y_i$, $x_i$) measures must furnish information about the relationship between X and Y. This is the situation for the example presented in Schafer [20, p. 54].

### 3.1. A near-linear regression solution
Equation (5) can be modified by introducing additional indicator variables for the missing X values, yielding for a bivariate situation

$$\begin{pmatrix} \mathbf{Y}_{o,o} \\ \mathbf{Y}_{x,m} \\ \mathbf{0}_{y,m} \end{pmatrix} = \begin{pmatrix} \mathbf{1}_{o,o} & \mathbf{X}_{o,o} \\ \mathbf{1}_{x,m} & \mathbf{0}_{x,m} \\ \mathbf{1}_{y,m} & \mathbf{X}_{y,m} \end{pmatrix} \begin{pmatrix} \beta_0 \\ \beta_X \end{pmatrix} + \sum_{m=1}^{Q} x_m \begin{pmatrix} \mathbf{0}_{o,o} \\ \mathbf{I}_{x,m} \\ \mathbf{0}_{y,m} \end{pmatrix} \beta_X - \sum_{m=1}^{P} y_m \begin{pmatrix} \mathbf{0}_{o,o} \\ \mathbf{0}_{x,m} \\ \mathbf{I}_{y,m} \end{pmatrix} + \varepsilon , \qquad (6)$$

where the subscript "o,o" denotes data that are complete in both X and Y, "x,m" denotes data that are incomplete in X and complete in Y, "y,m" denotes data that are incomplete in Y and complete in X, and the vectors of missing values to be estimated are $\mathbf{x}_m$ and $\mathbf{y}_m$. Of note here is that although $\mathbf{y}_m$ has a negative sign (it has been subtracted from both sides of the equation), $\mathbf{x}_m$ does not (it is being added because it has been removed from vector $\mathbf{X}$). Furthermore, equation (6) does not suffer from the usual drawbacks of conventional dummy variable adjustment [1, p. 9-11], because equation (6) contains a set of Q dummy variables, one for each missing value $x_m$, rather than a single dummy variable with Q ones and (n – Q) zeroes.

Even though equation (6) basically requires an OLS solution, it still needs to be estimated with a NLS routine (e.g., SAS PROC NLIN). The OLS solution based upon equation (6) for a bivariate case yields intercept ($b_0$) and slope ($b_1$) estimates based only on the complete data,

$$\mathbf{b}_o = (\mathbf{X}_{o,o}^T \mathbf{X}_{o,o})^{-1} \mathbf{X}_{o,o}^T \mathbf{Y}_{o,o} ,$$

and the missing value estimates

$$b_1\hat{x}_m = -b_0 1_{x,m} + y_{x,m} \Rightarrow \hat{x}_m = (-b_0 1_{x,m} + y_{x,m})/b_1$$
$$\hat{y}_m = b_0 1_{y,m} + b_1 x_{y,m},$$

are estimates that depend upon which variable is designated X and which is designated Y. The culprit is the correlation between X and Y, which enters into the equation as r in one instance, and 1/r in the other, resulting in $\mathbf{b}_{Y|X}$ and $\mathbf{b}_{X|Y}$—where Y|X denotes regressing Y on X and X|Y denotes regressing X on Y—failing to yield equivalent missing value estimates. A nonlinear procedure allows simultaneous minimization of the sum of squares quantity, creating a trade-off between the two linear regression solutions. Imputation calculations with such an implementation of equation (6) render the results reported in Table 2. But one serious drawback concerns degrees of freedom; in the example presented by Schafer [20, p. 54], the number of degrees of freedom already is 0.

For illustrative purposes, consider the first two variables in the SAS example dataset [19]; these are selected because no observation has a missing value for both of them. Calculation results for the preceding equations, together with the correct imputations, appear in Table 3. These results illustrate that the OLS solution furnishes upper and lower bounds for imputations, while rendering widely different standard error estimates. These results can be confirmed with SAS PROC MI. The estimated standard errors that it yields, based upon 10,000 MCMC iterations, suggest that the OLS Y|X solution yields the correct standard errors for the $y_m$s, and that the OLS X|Y solution yields the correct standard errors for the $x_m$s.

Table 3. A bivariate regression missing data imputation example using the SAS [19] sample data set.

| parameter | $X_1\|X_2$ estimates | | Simultaneous estimates | | $X_2\|X_1$ estimates | |
|---|---|---|---|---|---|---|
| | Point | Standard error | Point | Standard error | Point | Standard error |
| Observation #4, $X_1$ | 42.63 | **2.83** | 42.54 | 1.90 | 41.14 | 6.86 |
| Observation #20, $X_1$ | 46.02 | **2.78** | 46.01 | 1.87 | 45.84 | 6.86 |
| Observation #28, $X_1$ | 51.96 | **2.87** | 52.10 | 1.93 | 54.09 | 6.86 |
| Observation #6, $X_2$ | 9.38 | 1.66 | 9.42 | 0.27 | 9.75 | **0.78** |
| Observation #18, $X_2$ | 6.45 | 1.66 | 6.57 | 0.68 | 7.64 | **0.86** |
| Observation #23, $X_2$ | 10.44 | 1.66 | 10.45 | 0.59 | 10.52 | **0.77** |
| $a_{X_1\|X_2}$ | 79.44 | 4.31 | 79.75 | 2.93 | | |
| $a_{X_2\|X_1}$ | | | 22.56 | 3.14 | 21.58 | 1.42 |
| $b_{X_1\|X_2}$ | -3.08 | 0.40 | -3.11 | 0.27 | | |
| $b_{X_2\|X_1}$ | | | -0.26 | 0.07 | -0.23 | 0.03 |
| NOTE: simultaneous estimation results iteratively estimate Y|X and X|Y; subsequent equation (9) implements this method. | | | | | | |

### 3.2. Concatenation to obtain a solution

An ANCOVA specification can be written for each variable, regressing Y on X in one instance, and regressing X on Y in a second instance. Given equation (6), the two specifications may be written as follows:

$$\begin{pmatrix} \mathbf{Y}_{o,o} \\ \mathbf{Y}_{x,m} \\ \mathbf{0}_{y,m} \end{pmatrix} = \begin{pmatrix} \mathbf{1}_{o,o} & \mathbf{X}_{o,o} \\ \mathbf{1}_{x,m} & \mathbf{0}_{x,m} \\ \mathbf{1}_{y,m} & \mathbf{X}_{y,m} \end{pmatrix} \begin{pmatrix} \alpha_{Y|X} \\ \beta_{Y|X} \end{pmatrix} + \begin{pmatrix} \mathbf{0}_{o,o} \\ \mathbf{I}_{x,m} \\ \mathbf{0}_{y,m} \end{pmatrix} \mathbf{x}_m \beta_{Y|X} + \begin{pmatrix} \mathbf{0}_{o,o} \\ \mathbf{0}_{x,m} \\ -\mathbf{I}_{y,m} \end{pmatrix} \mathbf{y}_m + \varepsilon_{Y|X}, \text{ and} \qquad (7)$$

$$\begin{pmatrix} \mathbf{X}_{o,o} \\ \mathbf{0}_{x,m} \\ \mathbf{X}_{y,m} \end{pmatrix} = \begin{pmatrix} \mathbf{1}_{o,o} & \mathbf{Y}_{o,o} \\ \mathbf{1}_{x,m} & \mathbf{Y}_{x,m} \\ \mathbf{1}_{y,m} & \mathbf{0}_{y,m} \end{pmatrix} \begin{pmatrix} \alpha_{X|Y} \\ \beta_{X|Y} \end{pmatrix} + \begin{pmatrix} \mathbf{0}_{o,o} \\ -\mathbf{I}_{x,m} \\ \mathbf{0}_{y,m} \end{pmatrix} \mathbf{x}_m + \begin{pmatrix} \mathbf{0}_{o,o} \\ \mathbf{0}_{x,m} \\ \mathbf{I}_{y,m} \end{pmatrix} \mathbf{y}_m \beta_{X|Y} + \boldsymbol{\varepsilon}_{X|Y} , \quad (8)$$

where missingness is indexed by whether it occurs in X or Y, $\mathbf{I}_{x,m}$ and $\mathbf{I}_{y,m}$ respectively denote P-by-P and Q-by-Q identity matrices, and the intercept and slope parameters and error terms are subscripted according to whether Y is regressed on X, or X is regressed on Y. Resulting imputations are governed by the relationship between $X_o$ and $Y_o$. This formulation can be extended to any number of variables without loss of generality.

Data organized according to the appearance of terms in equations (7) and (8) can be concatenated for a simultaneous estimation of the parameters and missing values. This concatenation requires the creation and inclusion of two additional 0/1 indicator variables, one for each of the two equations. Once again ANCOVA as regression guides the specification. The resulting supra-ANCOVA may be written as

$$\begin{pmatrix} \mathbf{Y}_{y,o} \\ \mathbf{0}_{y,m} \\ \mathbf{X}_{x,o} \\ \mathbf{0}_{x,m} \end{pmatrix} = eq_1 \otimes \left[ \alpha_{Y|X} + \beta_{Y|X} \begin{pmatrix} \mathbf{X}_{x,o} \\ \mathbf{0}_{x,m} \end{pmatrix} + \beta_{Y|X} \begin{pmatrix} \mathbf{0}_{x,o} \\ \mathbf{I}_{x,m} \end{pmatrix} \mathbf{x}_m + \begin{pmatrix} \mathbf{0}_{y,o} \\ -\mathbf{I}_{y,m} \end{pmatrix} \mathbf{y}_m \right] +$$

$$eq_2 \otimes \left[ \alpha_{X|Y} + \beta_{X|Y} \begin{pmatrix} \mathbf{Y}_{y,o} \\ \mathbf{0}_{y,m} \end{pmatrix} + \beta_{X|Y} \begin{pmatrix} \mathbf{0}_{y,o} \\ \mathbf{I}_{y,m} \end{pmatrix} \mathbf{y}_m + \begin{pmatrix} \mathbf{0}_{x,o} \\ -\mathbf{I}_{x,m} \end{pmatrix} \mathbf{x}_m \right] + \begin{pmatrix} \boldsymbol{\varepsilon}_{y,o} \\ \mathbf{0}_{y,m} \\ \boldsymbol{\varepsilon}_{x,o} \\ \mathbf{0}_{x,m} \end{pmatrix} , \quad (9)$$

where $eq_k$ (k = 1, 2) denotes the two 2-by-1 binary 0/1 equation indicator variables, and $\otimes$ denotes Kronecker product. Estimation of this supra-equation requires iterative calculations, and can be executed efficiently with a NLS routine. Of note is that the indicator variables representing missing values in each variable include 0/1 values when the variable appears on the right-hand side of an equation, and 0/-1 values when the same variable originally appears on the left-hand side, but has been subtracted from both sides, of an equation.

Inspection of Table 2 reveals that this formulation of the problem solution renders exactly the results obtained by Schafer [20, p. 54]. By use of concatenation, the NLS regression routine is tricked into thinking that there are twice as many degrees of freedom than actually exist. Nevertheless, the estimate is an exact result (e.g., the error sum of squares is 0) because only two of ten observations have complete data. Furthermore, the correct solution for the difficult situation of a saddle point also is rendered by this formulation [see 12, p. 91], which can be used to identify maxima diverged to by a small perturbation in the starting correlation value, too. In this latter instance, the ANCOVA result deviates from the reported maxima result.

Next, consider a multiple variable case in which P values are missing for variable $X_1$, Q values are missing for variable $X_2$, and so on, but each observation has at least a measure on one of the variables. This is the situation for the example presented in Little and Rubin [11, p. 154], which includes five variables, with three having missing values. An ANCOVA specification can be written for each variable having missing values, regressing each X, in turn, on the remaining variables. As before, these sets of regression equations can be concatenated for simultaneous estimation using a NLS routine. Now a set of indicator variables is created for each variable having missing values. In the Little and Rubin example, three sets of indicator variables need to be created, and three sets of data need to be concatenated. Results from this super-ANCOVA appear in Table 2, and agree with reported results.[2] SAS code for implementing the EM algorithm in this context appears in Appendix A.

### 3.3. A univariate approximation for multivariate data

Efron [6] suggests a clever conversion of a set of multivariate data to a univariate missing data estimation problem by employing a two-way linear model; this most likely is a misspecified model that involves estimating a large number of parameters. The data are concatenated into a single vector, and a covariate matrix is constructed by creating (# rows – 1) and (# columns – 1) binary indicator variables. The resulting imputations are exactly those given by Theorem 4. Consequently,

> COROLLARY 1. By concatenating the entries in a n-by-p table of multivariate data into a single response variable, Y, resulting in missing values occurring only in this response variable, and then constructing (n-1) row and (p-1) column binary 0-1 indicator variables to be used as covariates, and, next, by replacing the missing values with zeroes and introducing a binary 0/-1 indicator variable covariate $-I_m$ for each missing value m, such that $I_m$ is 0 for all but missing value observation m and 1 for missing value observation m, the estimated regression coefficient $b_m$ is equivalent to the point estimate for a new observation, and as such furnishes EM algorithm imputations.

A severe weakness of this solution is that correlations, which almost certainly exist in multivariate data, are introduced into error terms and then ignored.

Efron notes that this is not necessarily a good imputation scheme (it most likely is misspecified). This solution for the Little and Rubin example reported in Table 4 illustrates his point[3]. Perhaps it yields reasonable imputation results when data are repeated measures, as in Efron's empirical example. But the SAS PROC MI variance estimates reported in Table 5 suggest that artificially creating covariates may not yield good imputation variance estimates even for data that appear to be normally distributed.

### 3.4. Variance estimates

The purpose of this section is to summarize explorations about estimates of variances for the multivariate case where data are missing in more than one variable, which are less straight forward. The concatenated nonlinear regression standard errors do not agree with those produced by multiple imputation using Markov chain Monte Carlo (MCMC) techniques, as they do when values are missing only in the response variable Y. The Little and Rubin [11, p. 154] example contains five variables, with the same four observations having missing values on the same three of these variables, and with one of these three variables also containing an additional three missing values; out of 65 possible data values, 50 are observed and 15 are missing. Both regression and multiple imputation results appear in Table 4. For this situation, the multiple imputation standard errors are a function of, in the following order (from a stepwise regression selection): the number of missing values for a given observation, the variance of the observed values for a given variable, and the imputed values. These three variates account for roughly 90% of the variance displayed by the multiple imputation standard errors.

---

[2] A typographical error may exist for the estimated mean of $X_4$, which is calculated here as 27.037 but reported as 27.047.
[3] The S-W probability for the pooled response variable is <0.0001, but the S-W probability for the residuals is 0.70.

Table 4. Multivariate normal imputations when more than one variable contains missing values: the Little and Rubin [11, p. 154] example dataset.

| Variable | Observation | unconstrained | | | | constrained | | unconstrained | |
|---|---|---|---|---|---|---|---|---|---|
| | | Equation (9) | | SAS PROC MI | | Equation (9) | | Efron's two-way linear model | |
| | | $\hat{y}_m$ | $s_{\hat{y}_m}$ | $\hat{y}_m$ | $s_{\hat{y}_m}$ | $\hat{y}_m$ | $s_{\hat{y}_m}$ | $\hat{y}_m$ | $s_{\hat{y}_m}$ |
| $X_1$ | 10 | 12.8907 | 196.1 | 12.8809 | 3.6786 | 12.8351 | 185.6 | 14.4444 | 16.8172 |
| | 11 | -0.4686 | 190.8 | -0.4561 | 3.6088 | 0 | 0 | 7.8944 | 16.8172 |
| | 12 | 9.9899 | 185.9 | 9.9999 | 3.5274 | 9.9935 | 177.2 | 15.6444 | 16.8172 |
| | 13 | 10.1054 | 179.3 | 10.0911 | 3.4224 | 10.0957 | 170.9 | 13.1944 | 16.8172 |
| $X_2$ | 10 | 65.8389 | 538.2 | 65.9448 | 9.4173 | 65.9877 | 509.5 | 53.4444 | 16.8172 |
| | 11 | 48.1963 | 523.5 | 48.1680 | 9.1891 | 46.9425 | 111.8 | 46.8944 | 16.8172 |
| | 12 | 68.0844 | 510.2 | 68.0420 | 8.9947 | 68.0745 | 486.2 | 54.6444 | 16.8172 |
| | 13 | 62.4285 | 492.2 | 62.4336 | 8.6654 | 62.4541 | 468.9 | 52.1944 | 16.8172 |
| $X_4$ | 7 | 0.8449 | 0.4569 | 0.8725 | 0.7982 | 0.8449 | 0.4335 | 49.9750 | 15.7310 |
| | 8 | 37.8695 | 0.4861 | 37.9041 | 0.8315 | 37.8695 | 0.4611 | 33.1750 | 15.7310 |
| | 9 | 19.8902 | 0.3400 | 19.8872 | 0.5886 | 19.8902 | 0.3225 | 43.3250 | 15.7310 |
| | 10 | 14.4591 | 366.3 | 14.3398 | 7.8093 | 14.3657 | 346.9 | 48.4972 | 17.1640 |
| | 11 | 20.7966 | 356.3 | 20.8345 | 7.6383 | 21.5841 | 148.4 | 41.9472 | 17.1640 |
| | 12 | 8.1828 | 347.2 | 8.2192 | 7.3767 | 8.1891 | 330.7 | 49.6972 | 17.1640 |
| | 13 | 15.4429 | 334.9 | 15.4612 | 7.1563 | 15.4269 | 318.9 | 47.2472 | 17.1640 |

Variance estimates were further explored by analyzing the SAS illustrative dataset that is furnished with discussion of SAS PROC MI [19, p. 133]. This example contains three variables—$X_1$, $X_2$ and $X_3$—with two of the same observations having missing values for $X_1$ and $X_3$, three of the same observations having missing values for $X_2$ and $X_3$, and the remaining missing values occurring only in $X_3$; out of 93 possible values, 78 are observed and 15 are missing. Both regression and multiple imputation results appear in Table 6. For this situation, the multiple imputation standard errors are a function of, in the following order (from a stepwise regression selection): the imputed values, and the number of missing values for a given observation. These two variates account for nearly all of the variance displayed by the multiple imputation standard errors.

These exploratory analyses indicate that analytical variances for the multiple imputations may well be available, but possibly are a function of the percentage and pattern of missing values across a battery of variables.

The bootstrap and Bayesian analysis offer two alternative procedures for variance estimation. Efron [6] describes the bootstrap as selecting n vectors from a set of n observation vectors by sampling with replacement, and then imputing the missing values for each sample. One advantage of this procedure is that it provides correct *large sample* estimates of standard errors when a model is misspecified (the two-way linear model frequently will be). Because different sets of missing values occur in each sample, 16,000 samples of size 22 were selected in order to ensure that each missing value was imputed roughly 10,000 times; as with SAS PROC MI, samples yielding values below a minimum (0 here) also were discarded. The Shapiro-Wilk (S-W) normality test statistics probability for the pooled observed values is 0.02, and for the two-way linear model residuals is 0.79. Nevertheless, for Efron's example (see Table 5), which can be considered a misspecified model, in some cases the bootstrap variance estimates are substantially less than the SAS PROC MI variance estimates, whereas the SAS PROC MI variances estimates always are at least noticeably less than the two-way linear model variance estimates. The bootstrap produces some unusually low variance estimates.

**Table 5**. Multivariate normal imputations when more than one variable contains missing values: the Mardia-Kent-Bibby example dataset in Efron [6, p. 464].

| Variable | Observation | equation (9) | | SAS PROC MI | | bootstrap[a] | | Bayesian[b] | |
|---|---|---|---|---|---|---|---|---|---|
| | | $\hat{y}_m$ | $s_{\hat{y}_m}$ | $\hat{y}_m$ | $s_{\hat{y}_m}$ | $\hat{y}_m$ | $s_{\hat{y}_m}$ | $\hat{y}_m$ | $s_{\hat{y}_m}$ |
| $X_1$ | 1 | 56.2066 | 9.1721 | 56.2872 | 7.8119 | 56.0440 | 4.9337 | 56.3096 | 8.1602 |
| | 4 | 47.9566 | 9.1721 | 48.1071 | 7.8130 | 47.8899 | 5.6498 | 47.6907 | 7.8177 |
| | 5 | 48.4566 | 9.1721 | 48.3281 | 7.8010 | 48.4061 | 5.7656 | 48.3576 | 8.0848 |
| | 6 | 49.9566 | 9.1721 | 49.9282 | 7.8740 | 49.7912 | 6.2077 | 50.2158 | 7.9448 |
| | 10 | 42.4566 | 9.1721 | 42.4494 | 7.8258 | 42.3080 | 5.2740 | 43.1281 | 8.3721 |
| | 11 | 39.5386 | 9.4455 | 39.4938 | 8.0793 | 38.4579 | 8.5772 | 39.6667 | 8.1990 |
| | 13 | 46.2053 | 9.4455 | 46.1134 | 8.1098 | 45.0144 | 7.9789 | 46.0618 | 8.4224 |
| | 14 | 45.2053 | 9.4455 | 45.1838 | 8.1521 | 44.0113 | 7.5730 | 45.0113 | 8.1725 |
| | 15 | 36.8719 | 9.4455 | 36.8037 | 8.1593 | 35.9970 | 6.3972 | 36.7116 | 8.3968 |
| | 21 | 20.4566 | 9.1721 | 20.6155 | 7.7557 | 20.2986 | 4.9312 | 20.5188 | 7.7712 |
| | 22 | 9.8719 | 9.4455 | 11.0954 | 6.6615 | 11.3022 | 8.7570 | 9.9758 | 8.2932 |
| $X_5$ | 3 | 58.9434 | 9.1721 | 59.0084 | 7.7260 | 58.9491 | 3.8756 | 59.0315 | 7.8953 |
| | 7 | 51.6934 | 9.1721 | 51.8166 | 7.7800 | 51.7384 | 4.6916 | 52.0758 | 7.9833 |
| | 8 | 46.9434 | 9.1721 | 46.8671 | 7.9598 | 46.8724 | 6.1322 | 46.7368 | 8.2842 |
| | 11 | 44.3281 | 9.4455 | 44.4390 | 8.1143 | 43.3540 | 8.1730 | 44.1594 | 8.1514 |
| | 12 | 48.1934 | 9.1721 | 48.2852 | 7.8123 | 48.1935 | 4.1424 | 48.0975 | 7.9460 |
| | 13 | 50.9947 | 9.4455 | 50.8717 | 8.0875 | 49.8699 | 7.5495 | 50.8233 | 8.3739 |
| | 14 | 49.9947 | 9.4455 | 49.9529 | 8.0293 | 48.8574 | 7.0725 | 50.2791 | 8.5207 |
| | 15 | 41.6614 | 9.4455 | 41.7093 | 8.1551 | 40.8671 | 5.9658 | 41.9486 | 8.5469 |
| | 17 | 37.6934 | 9.1721 | 37.7289 | 7.8176 | 37.5914 | 5.1018 | 37.8933 | 8.1998 |
| | 19 | 37.4434 | 9.1721 | 37.4915 | 7.8432 | 37.3092 | 6.7784 | 37.4017 | 7.9818 |
| | 22 | 14.6614 | 9.4455 | 15.2744 | 7.5559 | 16.2251 | 8.5820 | 14.7614 | 8.4393 |

[a] implemented with SAS PROC SURVEYSELECT.
[b] implemented with SAS PROC BGENMOD

A Bayesian analysis was completed using SAS PROC BGENMOD that posited normal priors for the imputations, with gamma priors for the variances of these normal priors. Because of high levels of serial correlations in the MCMC chain, only every 10th Gibbs sample value was retained; the first 2,000 selections were dropped as a burn-in sequence. Imputations reported in Table 5 are the averages of the resulting sample of 1,000 values. These thinned chains exhibit diagnostics for a good mixture: their time-series plots have no trend, their frequency distributions conform closely to a normal distribution, and 20 of the 22 have only trace serial correlation (two have lag-1 correlations of approximately 0.2).

Even though the four imputation estimators are nearly identical, none of the four variance estimators closely agree, although most of them are of the same order of magnitude. In Efron's example, equation (9) imputations always are accompanied by the largest standard errors, with the Bayesian imputations frequently accompanied by the 2nd largest. The bootstrap imputations frequently are accompanied by the smallest standard errors, with the SAS PROC MI imputations frequently accompanied by the 2nd smallest.

| Variable | Observation | Equation (9) $\hat{y}_m$ | Equation (9) $s_{\hat{y}_m}$ | SAS PROC MI $\hat{y}_m$ | SAS PROC MI $s_{\hat{y}_m}$ | bootstrap[a] $\hat{y}_m$ | bootstrap[a] $s_{\hat{y}_m}$ | Bayesian[b] $\hat{y}_m$ | Bayesian[b] $s_{\hat{y}_m}$ |
|---|---|---|---|---|---|---|---|---|---|
| $X_1$ | 4 | 41.3616 | 3.9332 | 42.4394 | 2.8472 | 44.979 | 1.0906 | 44.8852 | 4.8095 |
| | 20 | 45.9610 | 5.8598 | 46.2275 | 3.0015 | 46.197 | 2.8615 | 44.4865 | 5.0956 |
| | 28 | 52.1291 | 6.0461 | 52.2993 | 3.0663 | 48.506 | 2.6025 | 46.2250 | 5.1728 |
| $X_2$ | 6 | 9.4214 | 1.8602 | 9.7688 | 0.7996 | 10.596 | 0.6951 | 9.4110 | 1.5286 |
| | 18 | 6.6074 | 2.1145 | 7.6800 | 0.8897 | 10.156 | 0.8566 | 9.1234 | 1.6198 |
| | 23 | 10.4398 | 1.8310 | 10.5164 | 0.8030 | 10.743 | 0.7375 | 9.5446 | 1.5208 |
| $X_3$ | 3 | 170.6799 | 5.7283 | 170.357 | 10.5193 | 169.822 | 4.0047 | 168.7131 | 10.8533 |
| | 6 | 169.6093 | 6.5199 | 169.194 | 10.6017 | 169.659 | 28.9834 | 157.8428 | 21.8259 |
| | 10 | 172.2305 | 5.5869 | 171.993 | 10.2231 | 171.317 | 2.8331 | 170.6477 | 11.0300 |
| | 13 | 172.5714 | 5.5439 | 172.661 | 10.0909 | 172.072 | 2.1270 | 172.3324 | 11.1463 |
| | 18 | 164.7680 | 7.4900 | 163.350 | 11.7891 | 164.941 | 23.2533 | 157.0218 | 21.2332 |
| | 20 | 172.0624 | 8.4021 | 171.934 | 10.7925 | 171.942 | 34.6922 | 177.5195 | 14.7809 |
| | 23 | 171.3614 | 6.3984 | 171.347 | 10.4574 | 170.966 | 29.7830 | 158.5441 | 22.4279 |
| | 25 | 172.7761 | 5.6584 | 172.385 | 10.3858 | 171.366 | 3.3735 | 170.5800 | 10.6501 |
| | 28 | 168.7691 | 8.6998 | 168.562 | 11.2857 | 170.764 | 39.2247 | 175.6260 | 17.1730 |

Table 6. Multivariate normal imputations when more than one variable contains missing values: the SAS [19, p. 133] example dataset.

[a] implemented with SAS PROC SURVEYSELECT.
[b] implemented with SAS PROC BGENMOD

### 3.4.1. The bootstrap

The bootstrap with missing data, as described by Efron [6], may be ineffective when treating multivariate data as multivariate when they have a sizeable percentage of missing values distributed across multiple variables that cluster on a few observations [this is implied by discussion in 4], such as the Little and Rubin (11, p. 133) data (23% missing; 54% of observations with at least 1 missing value). Converting such data to a univariate problem with a two-way linear model for imputation purposes, *a la* Efron (reported in Table 5), yields standard errors that are inconsistent with regression results obtained with the two-way linear model specification, regardless of estimator. Imputation calculations exploit relationships between observed values, and in this data situation a given bootstrap sample can have most or all of its observations containing missing values.

The example 3-variable SAS dataset [19] has missing values for 32.3% of its observations, and 16.1% of all of its values missing. All three variables produce acceptable S-W diagnostic statistics. Here the bootstrap provides the smallest standard error estimates roughly 67% of the time, with the Bayesian estimates ranking largest or 2$^{nd}$ largest for all imputations (Table 6); rank orders are mixed across the estimators. Again, about 17,000 replicate samples were selected in order to obtain roughly 10,000 missing value estimates. Bootstrap results with these data for imputations obtained with the two-way linear model exhibit poor properties [e.g., pooled variables are non-normal, two-way linear model residuals produce a marginally acceptable S-W diagnostic statistic, not all average values are similar to those obtained with equation (9) and SAS PROC MI, and standard errors are relatively small].

### 3.4.2. Bayesian analysis

A Bayesian analysis of the example SAS data was completed with normal priors for the imputations, and with gamma priors for the variances of these normal priors. Because of high levels of serial correlations in the MCMC chain, only every 10$^{th}$ Gibbs sample value was retained for $X_1$ and $X_2$, and only every 300th value for $X_3$; the first 1,000 selections were dropped as a burn-in sequence. Imputations appearing in Table 6 are the averages of the resulting sample of 1,000 values. These thinned chains exhibit diagnostics for a good mixture: their time-series plots have no trend, their frequency distributions conform closely to a normal distribution, and 13 of the 15 have only trace serial correlation (two have lag-1 correlations of approximately 0.2).

### 3.4.3. Some implications about standard errors
Equation (9) is closely linked to the problem of predicting values for new observations. Accordingly, linear regression theory results in its accompanying standard errors being increased to account for the additional uncertainty (see Theorem 5), including the error from both a fitted model and the sampling of a new observation. Bayesian estimation shrinks this standard error by attaching a prior distribution to each imputation. Bootstrapping, which is complicated by the presence of missing values, provides a large sample standard error estimate for imputations in the presence of model misspecification. One advantage of both the Bayesian and bootstrap procedures is that they appear to provide increased variance estimates when multiple missing values occur for a given observation, in a multivariate treatment (see $X_3$ imputations for observations #6, #18, #20, #23, and #28 in Table 6).

## 4. EXTENSIONS TO SELECTED CONSTRAINED SITUATIONS
The EM algorithm is applied to compute many quantities besides missing values, including unknown variance components and latent variables. In these other contexts, an analyst may overlook important properties of the missing value estimates. One possible situation is the example presented by Little and Rubin [11, p. 154], in which the estimated conditional mean for missing value $x_{1,11}$ is -0.5. In some contexts values cannot be negative; this is true for the Little and Rubin example, too, because each of its first four variables is measured as a percentage of weight [see 5, p. 365]. This situation differs from including a minimum value for imputation purposes when executing SAS PROC MI.

By performing the ANCOVA estimation using NLS, constraints can be easily attached to individual parameters by placing bounds on their estimates. A minimum value of 0 constraint can be implemented either by placing a lower bound of 0 on a parameter estimate or replacing the parameter with an exponentiated one (e.g., $e^{\alpha_m}$). The Little and Rubin [11, p. 55] trivariate example furnishes a relatively simple illustration of this point. These data display certain features that suggest the four missing data values in each variable should have a mean of 2.5 and sum to 10. Imposing these two constraints results in estimates of approximately 2.5 for each of the missing values for variable $Y_1$, the set of values {2.46, 2.49, 2.51, 2.54} for variable $Y_2$, and the set of values {2.47, 2.49, 2.51, 2.53} for variable $Y_3$. These sets of values suggest the complete data pairwise correlations of $r(Y_1, Y_2) = 0.51284$, $r(Y_1, Y_3) = 0.51012$, and $r(Y_2, Y_3) = -0.47677$, which appear far more satisfactory than the respective reported values of 1, 1 and -1 produced by an *available-case* analysis. Next, the previous solution presented for the multiple variables illustration has been modified by restricting the missing value estimate for measure $x_{1,11}$ to be non-negative (i.e., a lower bound of 0). By doing so, the resulting estimate becomes 0; this value change is propagated through the covariance structure of the data, resulting in slight modifications to all missing value estimates for observation 11, and even slighter modifications to all missing values estimates for observation 10 (see Table 4).

Of note is that placing a constraint of 0 as a minimum for the Little and Rubin example prevents SAS PROC MI from properly executing. The minimum values specification for this SAS procedure ensure that MCMC-generated values are greater than the specified minimum value; when a simulated value is less than a specified minimum, the simulated sample chain is discarded, and a new simulated sample value is generated. But in a situation like the Little and Rubin example, virtually all simulated values will be less than 0, preventing the MCMC technique from successfully generating values.

### 4.1. Imposing a linear constraint
Sometimes the total of the missing data for a variable is known. In this situation, imputations can be constrained to sum to this known total. Accordingly,

> THEOREM 6. If the errors, $\boldsymbol{\varepsilon} = \mathbf{Y} - \mathbf{X}\boldsymbol{\beta}$, in equation (1) are distributed as MVN($\mathbf{0}, \mathbf{I}\sigma^2$) and $\sum_{k=1}^{m} \beta_m = T$, where T is a known total, then $\hat{\boldsymbol{\beta}}_{m-1} = \frac{1}{n_m}[(n_m \mathbf{I}_{m-1} - \mathbf{1}_{m-1}\mathbf{1}_{m-1}^T)\mathbf{X}_{m-1}\boldsymbol{\beta} - (\mathbf{X}_{n_m} - T)\mathbf{1}_{m-1}]$, where $\mathbf{X}_{n_m}$ is the row-vector of covariates for missing value $n_m$, is a MLE. And, $\hat{\boldsymbol{\beta}}_{n_m} = T - \mathbf{1}_{m-1}^T \hat{\boldsymbol{\beta}}_{m-1}$.

> PF: For observation $n_m$ in the log-likelihood equation, $T - \mathbf{1}_{m-1}^T \hat{\boldsymbol{\beta}}_{m-1}$ replaces $\hat{\boldsymbol{\beta}}_{n_m}$. As with Theorem 1, the solution results from solving $\frac{\partial \text{LN}(L)}{\partial \boldsymbol{\beta}_{m-1}} = 0$.
>
> $\frac{\partial^2 \text{LN}(L)}{\partial \beta_{m-1}^2} = -\frac{2}{\sigma^2} \forall m-1$
>
> $\therefore \hat{\boldsymbol{\beta}}_{m-1} = \frac{1}{n_m}[(n_m \mathbf{I}_{m-1} - \mathbf{1}_{m-1}\mathbf{1}_{m-1}^T)\mathbf{X}_{m-1}\boldsymbol{\beta} - (\mathbf{X}_{n_m} - T)\mathbf{1}_{m-1}]$ is MLE □

The following corollary relates this result to the mean of the missing values:

> COROLLARY 2. If no covariates are present, then $\mathbf{X}_{m-1} = \mathbf{1}_{m-1}$, and $\hat{\boldsymbol{\beta}}_m = T/n_m$.
>
> PF: Substitute $\mathbf{1}_{m-1}$ for $\mathbf{X}_{m-1}$ in Theorem 6.

One variant of this constraint, when a total for the missing values is unknown, is to preserve the mean of the known data by setting $T = n_m \bar{y}_o$.

### 4.2. Imposing a nonlinear constraint when 0 is the minimum
NLS supports implementation of nonlinearly constrained estimation. The Little and Rubin [11] example, whose results appear in Table 4, can have the relatively simple 0 constraint imposed in two ways: include $0 < \beta_{1,11}$ as a lower bound on the missing value estimate; or, estimate the parameter $\alpha_{11}$ for $\beta_{1,11} = e^{\alpha_{11}}$. Either of these constraints yields the results reported in Table 4, except for slight rounding error differences.

Theorem 6 allows negative imputations. When missing values have a minimum of 0, implementing a total constraint requires a binomial type of specification: $\beta_{m-1} = \frac{e^{\alpha_{m-1}}}{1 + \sum_{k=1}^{n_{m-1}} e^{\alpha_k}} T$ and

$$\beta_{n_m} = \frac{1}{1+\sum_{k=1}^{n_m-1} e^{\alpha_k}} T.$$ For the Efron data example, attaching this percentage-of-totals constraint (missing value totals are 401 for $X_1$ and 434 for $X_5$) reduces by roughly 11% the root mean squared error (RMSE) between the imputations and their corresponding observed values (Table 7).

| variable | Missing value observation | Actual value | Nonlinear constraint imputation | Standard error | Efron's imputation |
|---|---|---|---|---|---|
| $X_1$ | 1 | 63 | 52.94 | 2.6439 | 56.21 |
|  | 4 | 44 | 44.69 | 2.2319 | 47.96 |
|  | 5 | 42 | 45.19 | 2.2569 | 48.46 |
|  | 6 | 31 | 46.69 | 2.3318 | 49.96 |
|  | 10 | 36 | 39.19 | 1.9572 | 42.46 |
|  | 11 | 56 | 35.02 | 1.7487 | 39.54 |
|  | 13 | 40 | 41.69 | 2.0816 | 46.21 |
|  | 14 | 23 | 40.69 | 2.0317 | 45.21 |
|  | 15 | 49 | 32.35 | 1.6155 | 36.87 |
|  | 21 | 12 | 17.19 | 0.8586 | 20.46 |
|  | 22 | 5 | 5.35 | 0.2672 | 9.87 |
| $X_5$ | 3 | 68 | 55.04 | 2.6422 | 58.94 |
|  | 7 | 46 | 47.79 | 2.2942 | 51.69 |
|  | 8 | 64 | 43.04 | 2.0662 | 46.94 |
|  | 11 | 35 | 39.30 | 1.8864 | 44.33 |
|  | 12 | 33 | 44.29 | 2.1262 | 48.19 |
|  | 13 | 25 | 45.96 | 2.2064 | 50.99 |
|  | 14 | 44 | 44.96 | 2.1584 | 49.99 |
|  | 15 | 39 | 36.63 | 1.7584 | 41.66 |
|  | 17 | 43 | 33.79 | 1.6221 | 37.69 |
|  | 19 | 17 | 33.54 | 1.6101 | 37.44 |
|  | 22 | 20 | 9.63 | 0.4623 | 14.66 |

Table 7. Two-way linear model imputations for the Efron [6] example data

The Little and Rubin [11, p. 154] example data imputations reported in Table 4 yield the following RMSEs:

| Equation (9) | SAS PROC MI | Equation (9) with 0 minimum constraint | Two-way linear model | Equation (9) with total constraint |
|---|---|---|---|---|
| 61.93 | 62.22 | 59.79 | 421.91 | 45.19 |

These results reflect that the imputations obtained with an unconstrained equation (9) and SAS PROC MI are similar. Interestingly, imposing a minimum 0 constraint with NLS imputations improves the correspondence with the observed values. As noted previously, the two-way linear model proposed by Efron is artificial, and misspecified, and provides very poor imputations. Finally, the 3-totals constraint (43 for $X_1$, 221 for $X_2$, and 156 for $X_3$) reduces the RMSE by roughly 25% from the first three estimators. Furthermore, in all but two cases, the totals constraint substantially shrinks the standard errors from those produced by SAS PROC MI. The standard error for the 0-constrained imputation remains at 0.

## 5. DISCUSSION

Six theorems are presented that help to furnish insights into and simplify computational demands of the EM algorithm solution, at least for selected but common data cases, relating this solution based upon the linear model with a normal probability distribution for its error term to well-known standard regression theory for predicting values of new observations. These theorems furnish analytical means and variances for the case of missingness in the response variable only. Extensions to missingness in a battery of variables are achieved with nonlinear regression. One result that merits closer scrutiny is the failure to obtain maxima estimates that agree with those reported by McLachlan and Krishnan [12, p. 91; see Table 2], which reflects the utilization of a different optimization criterion. Future research needs to establish the cases in which both the EM algorithm and the ANCOVA specification yield the same results.

Standard errors provided by both bootstrapping and Bayesian solutions suggest that least squares solutions may overestimate these quantities. Although the bootstrap is expected to provide large sample standard errors when a model is misspecified, selected analyses with two-way linear model imputations presented by Efron [6] indicate that it apparently cannot compensate for extreme misspecification. Including constraints when totals are known appears to provide improved imputations with smaller standard errors.

# APPENDIX A
## SAS CODE [see 18] FOR IMPUTING BOTH X AND Y VALUES

One of the Little and Rubin [11, p. 154] datasets is employed here. The first part of the SAS code inputs these data.

```
DATA STEP1; INPUT X1 X2 X3 X4 X5;
CARDS;
 7 26  6 60  78.5
 1 29 15 52  74.3
11 56  8 20 104.3
11 31  8 47  87.6
 7 52  6 33  95.9
11 55  9 22 109.2
 3 71 17  . 102.7
 1 31 22  .  72.5
 2 54 18  .  93.1
 .  .  4  . 115.9
 .  . 23  .  83.8
 .  .  9  . 113.3
 .  .  8  . 109.4
;
RUN;
```

The next SAS code creates a binary indicator variable for each missing data value, according to its position in the data sequence, and replaces missing data in the dataset with a zero.

```
DATA STEP1(REPLACE=YES); SET STEP1;
IF _N_=7  THEN I7= 1; ELSE I7= 0;
IF _N_=8  THEN I8= 1; ELSE I8= 0;
IF _N_=9  THEN I9= 1; ELSE I9= 0;
IF _N_=10 THEN I10=1; ELSE I10=0;
IF _N_=11 THEN I11=1; ELSE I11=0;
IF _N_=12 THEN I12=1; ELSE I12=0;
IF _N_=13 THEN I13=1; ELSE I13=0;
IF X1='.' THEN X1=0;
IF X2='.' THEN X2=0;
IF X4='.' THEN X4=0;
RUN;
```

The next three sets of SAS code respectively create the parts, one for each variable with missing values, needed for the concatenated dataset used to estimate the missing values. Then these three parts are concatenated.

```
DATA STEP2A; SET STEP1;
EQ1=1; EQ2=0; EQ4=0;
Y=X1; Z1=X3; Z2=X5; Z3=X2; Z4=X4;
DROP X1-X5;
RUN;

DATA STEP2B; SET STEP1;
EQ1=0; EQ2=1; EQ4=0;
Y=X2; Z1=X3; Z2=X5; Z3=X1; Z4=X4;
DROP X1-X5;
RUN;
```

```
        DATA STEP2C; SET STEP1;
        EQ1=0; EQ2=0; EQ4=1;
        Y=X4; Z1=X3; Z2=X5; Z3=X1; Z4=X2;
        DROP X1-X5;
        RUN;

        DATA STEP2; SET STEP2A STEP2B STEP2C; RUN;
```

The following SAS code implements the ANCOVA needed to calculate the imputations [see equation (9)]. Each missing value has an indicator and a regression parameter associated with it (e.g., I10 and M10 for observation 10 in variable X1):

```
        PROC NLIN NOITPRINT MAXITER=500 METHOD=MARQUARDT;
         PARMS A1=0 A2=1 A3=1 A4=1 A5=1
               B1=0 B2=1 B3=1 B4=1 B5=1
                C1=0 C2=1 C3=1 C4=1 C5=1
               M10= 6 M11= 6 M12= 6 M13= 6
               N10=45 N11=45 N12=45 N13=45
               P7 =39 P8 =39 P9 =39 P10=39 P11=39 P12=39 P13=39;
        MODEL Y = EQ1*(A1 + A2*Z1 + A3*Z2
                  + A4*(Z3 + N10*I10 + N11*I11 + N12*I12 +N13*I13)
                  + A5*(Z4 + P7*I7 + P8*I8 + P9*I9 + P10*I10
                  + P11*I11 + P12*I12 +P13*I13)
                     -  (M10*I10 + M11*I11 + M12*I12 +M13*I13) ) +
                EQ2*(B1 + B2*Z1 + B3*Z2
                  + B4*(Z3 + M10*I10 + M11*I11 + M12*I12 +M13*I13)
                  + B5*(Z4 + P7*I7 + P8*I8 + P9*I9 + P10*I10
                  + P11*I11 + P12*I12 +P13*I13)
                  - (N10*I10 + N11*I11 + N12*I12 +N13*I13) ) +
                EQ4*(C1 + C2*Z1 + C3*Z2
                  + C4*(Z3 + M10*I10 + M11*I11 + M12*I12 +M13*I13)
                  + C5*(Z4 + N10*I10 + N11*I11 + N12*I12 +N13*I13)
                     -  (P7*I7 + P8*I8 + P9*I9 + P10*I10 + P11*I11
                    + P12*I12 +P13*I13) );
        RUN;
```

This SAS code pools the three estimation equations─differentiating their specifications with the binary indicator variables EQ1, EQ2 and EQ4─and has the missing value regression parameters on both sides of the model equation (on the left-hand side with the indicator variable is subtracted, and on the right-hand side when the indicator variable is added). Recall that each missing value has been replaced with a 0.

If a missing value imputation is to be constrained, a BOUNDS statement can be included. Regression parameters can be initialized with values from complete case regressions, and imputations can be initialized with corresponding means of the know variable values.